\documentclass[10pt,twocolumn]{article}

\usepackage[a4paper,margin=0.75in]{geometry}
\usepackage{amsmath,amssymb,amsfonts}
\usepackage{graphicx}
\usepackage{booktabs}
\usepackage{cite}
\usepackage{url}
\usepackage{hyperref}

\hypersetup{
    colorlinks=true,
    linkcolor=blue,
    citecolor=blue,
    urlcolor=blue
}

\title{\textbf{Comprehensive Review of Advances and Challenges in Next Generation Wireless Networks: From Novel Hardware Technologies to Learning Based Resource Allocation in 6G}}

\author{
\begin{tabular}{c}
\textbf{Armin Farhadi} \\
School of Electrical and Computer Engineering \\
College of Engineering, University of Tehran
\end{tabular}
\and
\begin{tabular}{c}
\textbf{Ali Olfat} \\
School of Electrical and Computer Engineering \\
College of Engineering, University of Tehran
\end{tabular}
}

\date{}

\begin{document}

\maketitle

% -------------------------------------------------
%                  Abstract
% -------------------------------------------------
\begin{abstract}
In modern wireless communication systems, there is a rapidly increasing demand for connectivity to wireless networks. Devices such as internet of things (IoT) devices, connected vehicles, smartphones, surveillance systems, and various other applications contribute significantly to this demand. Consequently, next-generation wireless systems must be capable of handling this enormous volume of devices and traffic. In recent years, several technologies have been introduced to address these challenges, including reconfigurable intelligent surfaces (RIS), integrated sensing and communication (ISAC), advanced antenna and intelligent surface technologies, and novel multiple access (MA) techniques. Furthermore, due to the limited resources available in communication systems, efficient resource allocation strategies are essential to support complex and high-dimensional optimization problems. In addition, modern communication systems are required to optimize resources within strict time constraints. Therefore, resource allocation solutions must be intelligent and computationally efficient. Conventional optimization techniques, such as convex optimization, are often inadequate for addressing these requirements. To overcome these limitations, novel resource allocation algorithms based on learning methods have been developed. In this paper, we comprehensively investigate advanced communication technologies alongside modern resource allocation optimization methods and algorithms based on machine learning techniques. Subsequently, current challenges of wireless networks are analyzed. Finally, open research challenges are identified.
\end{abstract}

% -------------------------------------------------
%                  Keywords
% -------------------------------------------------

% -------------------------------------------------
%                Introduction
% -------------------------------------------------
\section{Introduction}

Wireless communication systems are moving toward sixth-generation (6G) networks to support very high data rates, low latency, reliable communication, and intelligent services. New applications such as internet of things (IoT), autonomous systems, immersive technologies, digital twins, and integrated sensing and communication (ISAC) require significant changes in network design, signal processing, and resource allocation strategies. In this situation, designing efficient resource allocation methods becomes very important, since the network must support different types of services with diverse and sometimes conflicting requirements~\cite{ref1,ref2,ref3}.

Next-generation wireless networks, especially 6G systems, are expected to provide not only high-quality communication services but also accurate sensing capabilities. In this context, ISAC has emerged as a promising technology that combines sensing and communication functions into a unified framework~\cite{ref1}. In general, sensing aims to extract useful information from noisy signals, while communication focuses on delivering information over noisy channels. ISAC integrates these two functions to improve spectral efficiency, energy efficiency, and hardware utilization, while also enabling new applications in which sensing and communication support each other. Unlike conventional systems, where these two functions compete for resources, ISAC adopts a joint design approach to achieve better overall performance. Early studies have shown that sensing and communication share many common features, especially in beamforming and antenna array processing~\cite{ref1}. With the development of advanced technologies such as massive multiple-input multiple-output (MIMO) and millimeter-wave (mmWave) communications, the integration of sensing and communication has become more practical, since both rely on large antenna arrays and high-frequency bands~\cite{ref1}.

At the same time, modern wireless systems are moving toward a new paradigm in which not only the transmitter and receiver but also the propagation environment can be controlled. This concept is enabled by reconfigurable intelligent surfaces (RIS)~\cite{ref1,ref2,ref3,ref4,ref5,ref6}, which consist of passive elements capable of adjusting the phase of incident signals. Although conventional RIS can improve system performance, it has some limitations, such as limited control over signal properties and restricted coverage. To overcome these challenges, advanced structures have been introduced. For example, beyond-diagonal RIS (BD-RIS) allows interactions between elements, enabling control over both amplitude and phase of signals~\cite{ref2}. Similarly, simultaneously transmitting and reflecting RIS (STAR-RIS)~\cite{ref3} can reflect and transmit signals at the same time, providing full-space coverage and more flexible signal control~\cite{ref3}. These technologies significantly improve system performance and offer higher flexibility compared to traditional designs.

In addition to intelligent surfaces, advanced antenna technologies also play an important role in future wireless networks~\cite{ref7,ref8}. New antenna architectures aim to improve spectral efficiency, energy efficiency, and system reliability~\cite{ref7,ref8}. Flexible antenna systems, such as fluid antenna systems, can adapt their positions and configurations to improve channel conditions. Flexible intelligent metasurfaces (FIM) can also dynamically adjust their structure to enhance signal propagation. These technologies are particularly useful in high-frequency communication scenarios, where channel conditions vary rapidly~\cite{ref7,ref8}. The combination of advanced antennas and intelligent surfaces creates highly adaptive wireless environments that support efficient resource allocation and improved system performance~\cite{ref7,ref8}.

Furthermore, future wireless networks must support a large number of users while maintaining high spectral efficiency. In this regard, non-orthogonal multiple access (NOMA) has been proposed as an effective solution~\cite{ref9}. Unlike conventional orthogonal schemes, NOMA allows multiple users to share the same resources by managing interference. Sparse code multiple access (SCMA) further improves performance by using sparse codebooks and efficient multi-user detection techniques~\cite{ref1,ref9}.

To address the limitations related to the conventional convex optimization-based resource allocation, learning-based approaches have been introduced for wireless resource allocation~\cite{ref1}. In particular, deep reinforcement learning (DRL) has shown strong potential in solving complex optimization problems such as power allocation, beamforming design, and antenna configuration~\cite{ref6}. DRL can learn optimal strategies through interaction with the environment without requiring accurate mathematical models. However, conventional DRL methods may suffer from limited generalization when system conditions change~\cite{ref7}. To overcome this issue, meta-learning has been proposed as an advanced approach that enables fast adaptation to new environments by learning from multiple tasks. This improves robustness and reduces the need for repeated training~\cite{ref2}.

% -------------------------------------------------
%                Related Works
% -------------------------------------------------
\subsection{Related Works}

ISAC has been widely studied in recent years, particularly in the context of joint radar and communication system design and spectrum sharing frameworks~\cite{ref1,ref10}. Many existing works have focused on developing beamforming and resource allocation strategies under different system configurations. For instance, ISAC-based frameworks have been investigated in~\cite{ref1}, while transmit beamforming optimization for massive MIMO-based ISAC systems has been studied in~\cite{ref6}. In addition, robust and secure transmission under QoS constraints has been analyzed in~\cite{ref11}, and dynamic power allocation for vehicular ISAC networks has been considered in~\cite{ref12}. Although these studies provide useful insights into the tradeoff between sensing and communication, most of them rely on conventional architectures and classical optimization techniques, and they consider only limited recent technologies.

RIS-assisted systems have also attracted significant attention due to their ability to improve spectral and energy efficiency. Early works mainly focused on joint active and passive beamforming design, transmit power minimization, and sum-rate maximization in RIS- and STAR-RIS-assisted networks~\cite{ref13,ref14}. Further studies investigated cooperative beamforming in multi-RIS systems and advanced transmission and reflection models~\cite{ref13,ref14}. In this direction,~\cite{ref1} studied a STAR-BD-RIS-assisted SCMA-based ISAC system operating in the THz band and formulated a joint beamforming and resource allocation problem under QoS, coupled phase-shift, and channel state information (CSI) constraints, where meta-learning-based methods were applied to obtain efficient solutions.

SCMA has been widely studied as an effective approach for improving spectral efficiency and supporting massive connectivity. Existing works have investigated joint power and codebook design~\cite{ref5,ref9}, and intelligent surface-assisted uplink and downlink SCMA systems~\cite{ref15}. Energy-efficient and secure transmission strategies under both perfect and imperfect CSI conditions have also been examined~\cite{ref16}.

Recently, flexible antenna and intelligent metasurface technologies, such as flexible intelligent metasurfaces, have been introduced as promising solutions for dynamic wireless environments. Prior works have investigated joint optimization of transmit power, phase shifts, and electromagnetic responses in FIM-assisted systems~\cite{ref7,ref8}.

To address resource allocation-related challenges, learning-based approaches have been introduced for wireless resource allocation. In particular, DRL has been widely applied to solve complex optimization problems such as beamforming and power control~\cite{ref17,ref18,ref19,ref20}. Algorithms such as DDPG enable model-free optimization and can handle continuous decision variables. To further improve adaptability under dynamic network conditions, meta-learning-assisted DRL frameworks have been proposed for new generation wireless systems~\cite{ref1,ref2,ref3,ref4,ref6,ref7,ref8}, demonstrating faster convergence and improved robustness compared to conventional DRL methods.

% -------------------------------------------------
%               Key Contributions
% -------------------------------------------------
\subsection{Key Contributions}

This paper provides a comprehensive review of recent advances in wireless communication systems, with a focus on integrated technologies and resource allocation methods for next-generation networks. The main contributions are summarized as follows:

\begin{itemize}

\item We present a structured overview of ISAC systems, highlighting their fundamental principles and system architectures.

\item We investigate RIS-assisted wireless systems, including conventional RIS, STAR-RIS, and advanced architectures such as BD-RIS and STAR-BD-RIS. The role of these technologies in enhancing spectral efficiency, energy efficiency, and spatial control is discussed, with emphasis on their integration into complex wireless environments.

\item We provide a detailed discussion of advanced multiple access schemes, including SCMA, analyzing their principles, advantages, and limitations in terms of spectral efficiency, interference management, and support for massive connectivity.

\item We review flexible antenna and intelligent metasurface technologies, including FIM systems, and highlight their ability to dynamically adapt antenna structures, improve channel conditions, and enhance beamforming and resource allocation performance.

\item We analyze both traditional and learning-based resource allocation methods. Classical optimization techniques are first discussed, followed by a review of deep reinforcement learning approaches such as DDPG. Furthermore, we highlight the role of meta-learning in improving adaptability and convergence in dynamic wireless environments.

\item Finally, we identify key challenges and open research problems in the joint integration of ISAC, RIS-based architectures, advanced multiple access schemes, advanced antenna technologies, and intelligent learning-based optimization methods, including scalability, real-time implementation, and the need for unified optimization frameworks.

\end{itemize}

% -------------------------------------------------
%            Organization of the Paper
% -------------------------------------------------
\subsection{Organization of the Paper}

The rest of this paper is organized as follows. Section II provides a comprehensive review of ISAC systems, including their basic principles and system models. Section III discusses RIS-assisted wireless systems, covering conventional RIS, STAR-RIS, and advanced architectures such as BD-RIS and STAR-BD-RIS, along with their role in improving system performance. Section IV presents advanced multiple access techniques, including SCMA, and reviews their applications in supporting massive connectivity and improving spectral efficiency. Section V focuses on flexible antenna and intelligent metasurface technologies, including FIM systems, and their impact on adaptive wireless environments. Section VI reviews traditional and learning-based resource allocation methods, including classical optimization techniques, deep reinforcement learning, and meta-learning approaches. Section VII discusses open research challenges and future directions. Finally, Section VIII concludes the paper.

% -------------------------------------------------
%                      ISAC
% -------------------------------------------------
\section{ISAC}

\begin{figure}[!t]
\centering
\includegraphics[width=1\columnwidth]{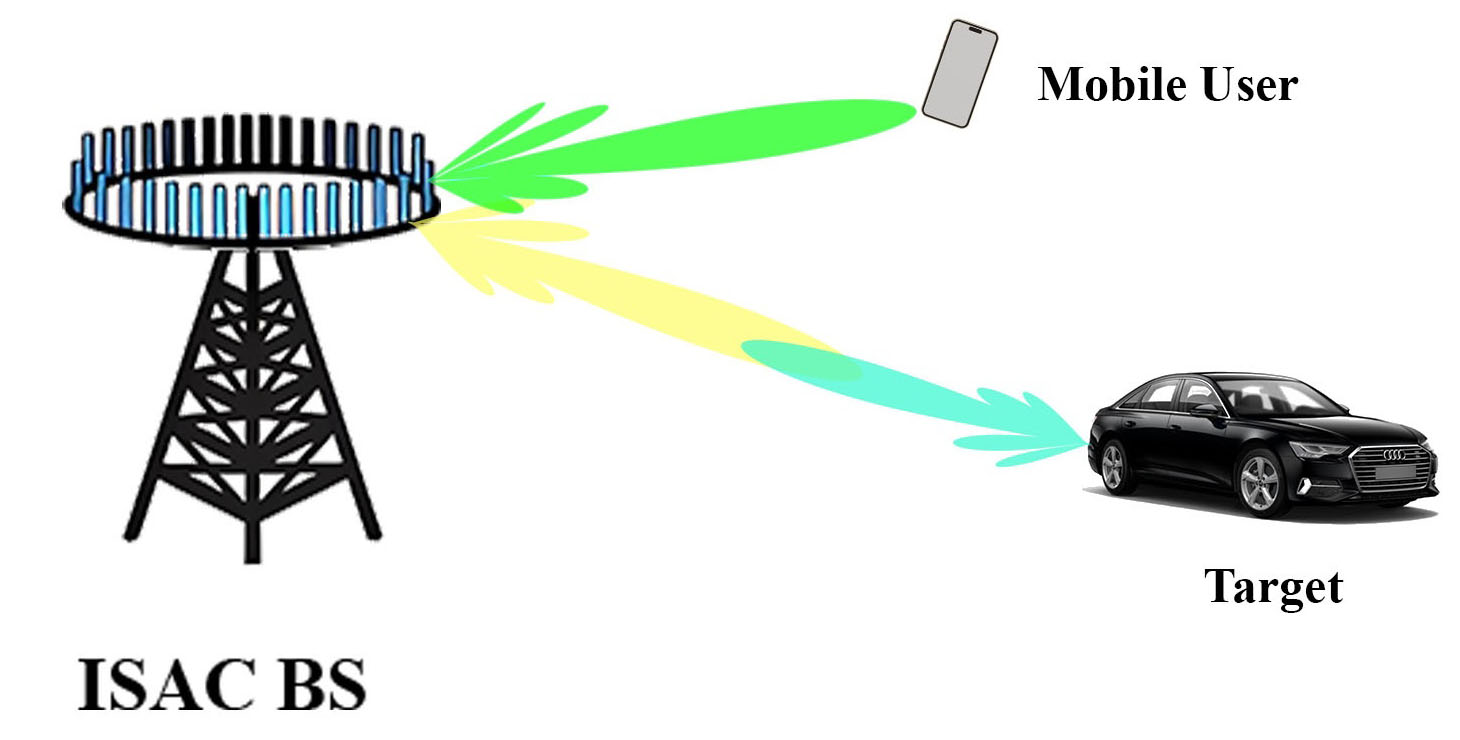}
\caption{ISAC-based communication networks.}
\label{fig:isac_network}
\end{figure}

This section provides a comprehensive review of ISAC systems, focusing on their basic principles and system models. The main idea of ISAC is to enable sensing and communication functionalities within a unified wireless framework~\cite{ref1}, as illustrated in Fig.~\ref{fig:isac_network}. In conventional wireless systems, sensing and communication are usually designed and operated separately. However, in ISAC systems, both functions are jointly considered to improve the efficiency of spectrum usage, energy consumption, and hardware utilization.

The basic principle of ISAC relies on sharing wireless resources such as spectrum, waveforms, and antenna arrays between sensing and communication tasks. This shared design allows the transmitted signals to carry communication information while also being used for sensing the surrounding environment. As a result, ISAC systems can simultaneously estimate target parameters such as location, velocity, and range, while also delivering data to communication users~\cite{ref1}. From a system modeling perspective, ISAC frameworks typically consist of a transmitter, multiple communication users, and sensing targets. The transmitter, often equipped with multiple antennas, generates a unified waveform that serves both sensing and communication purposes~\cite{ref1}. The received signals at communication users are used for data detection, while the reflected or echoed signals from targets are used for sensing parameter estimation.

Depending on the design approach, ISAC system models can be classified into different categories. In communication-centric models, sensing functionality is embedded into existing communication waveforms. In radar-centric models, communication signals are integrated into radar transmission frameworks. In fully integrated models, a common waveform and joint signal processing framework are designed to support both sensing and communication simultaneously~\cite{ref1}.

Mathematically, the transmitted signal in ISAC systems can be expressed as~\cite{ref1}

\begin{equation}
\mathbf{x}=\mathbf{Ws},
\label{eq:isac_tx}
\end{equation}

where $\mathbf{W}$ denotes the beamforming matrix and $\mathbf{s}$ contains the information-bearing symbols for both sensing and communication tasks.

The received signal at the communication users is given by~\cite{ref1}

\begin{equation}
\mathbf{y}=\mathbf{Hx}+\mathbf{n},
\label{eq:isac_rx}
\end{equation}

where $\mathbf{H}$ represents the communication channel matrix and $\mathbf{n}$ denotes additive noise.

For sensing functionality, the received radar echo signal can be modeled as~\cite{ref1}

\begin{equation}
{{\mathbf{Y}}_{\text{radar}}}
=
\sum\limits_{u=1}^{U}
{{{a}_{u}}}
{{\alpha }_{r}}({{\phi }_{u}})
\alpha _{t}^{H}({{\phi }_{u}})
\mathbf{x}
+
\mathbf{N},
\label{eq:radar_echo}
\end{equation}

where ${{a}_{u}}$ is the complex path gain of the $u$-th target, and ${{\alpha }_{t}}(\phi )$ and ${{\alpha }_{r}}(\phi )$ are the transmit and receive steering vectors, respectively.

The main design challenge lies in the tradeoff between sensing and communication performance. Improving sensing accuracy may reduce communication quality and vice versa. Therefore, beamforming design and resource allocation must be carefully optimized to balance both objectives. For sensing performance evaluation, the beam pattern error is defined as~\cite{ref1}

\begin{equation}
{{\epsilon }_{\text{Beam}}}(\mathbf{R})
=
\frac{1}{L}
\sum\limits_{l=1}^{L}
{{{\left|
P({{\theta }_{l}})
-
d({{\theta }_{l}})
\right|}^{2}}},
\label{eq:beam_error}
\end{equation}

where $\mathbf{R}$ is the transmit covariance matrix, $P({{\theta }_{l}})$ is the generated beampattern power, and $d({{\theta }_{l}})$ is the desired beam pattern response~\cite{ref1}.

Overall, ISAC system models provide the foundation for analyzing and designing future wireless networks that require both high-quality communication and accurate environmental sensing capabilities.

% -------------------------------------------------
%         RIS-Assisted Wireless Systems
% -------------------------------------------------
\section{RIS-Assisted Wireless Systems}

RIS technology has emerged as a promising solution to enhance the performance of future wireless communication systems. The main idea of RIS is to control and reconfigure the wireless propagation environment by adjusting the electromagnetic response of a large number of passive reflecting elements. This enables improved signal strength, better coverage, and enhanced spectral and energy efficiency without requiring additional active radio frequency chains~\cite{ref2,ref3,ref4}.

In conventional RIS architectures, each reflecting element independently adjusts the phase of the incident signal. This simple structure allows basic beamforming capabilities and has been widely studied for improving communication performance~\cite{ref4}. However, conventional RIS is limited in terms of flexibility, as it mainly supports reflection-only operation and cannot fully exploit advanced signal manipulation capabilities~\cite{ref3}.

To overcome these limitations, STAR-RIS has been introduced. Unlike conventional RIS, STAR-RIS~\cite{ref3,ref22} can simultaneously reflect and transmit incident signals, enabling full-space coverage and more flexible control of electromagnetic waves. This feature significantly improves system performance, especially in scenarios where users are located on both sides of the surface.

Further enhancements have led to BD-RIS~\cite{ref2}, which introduces inter-element coupling and more general scattering behavior. In BD-RIS, the scattering matrix is no longer restricted to a diagonal structure, allowing more degrees of freedom in signal processing and beamforming design. This results in improved system performance in terms of spectral efficiency, coverage, and interference management.

The dual-sector BD-RIS framework can be modeled using two complex matrices denoted as ${{\mathbf{\Phi }}^{s}}\in {{\mathbb{C}}^{F\times F}}$ for each sector $s\in \mathcal{S}$. These matrices represent the reflection and transmission behavior of each sector and are extracted from the overall scattering matrix $\mathbf{\Phi }\in {{\mathbb{C}}^{2F\times 2F}}$, which models the full impedance-based structure of the surface. Specifically, each sub-matrix is given by

\begin{equation}
{{\mathbf{\Phi }}^{s}}={{[\mathbf{\Phi }]}_{F+1:2F,1:F}}.
\label{eq:phi_submatrix}
\end{equation}

The resulting matrices satisfy a unitary power-preserving constraint expressed as

\begin{equation}
\sum\limits_{s}{{{({{\mathbf{\Phi }}^{s}})}^{H}}}{{\mathbf{\Phi }}^{s}}={{\mathbf{I}}_{F}},
\label{eq:unitary_constraint}
\end{equation}

which ensures energy conservation across different sectors~\cite{ref2,ref7,ref8}.

\begin{figure}[!t]
\centering
\includegraphics[width=0.7\columnwidth]{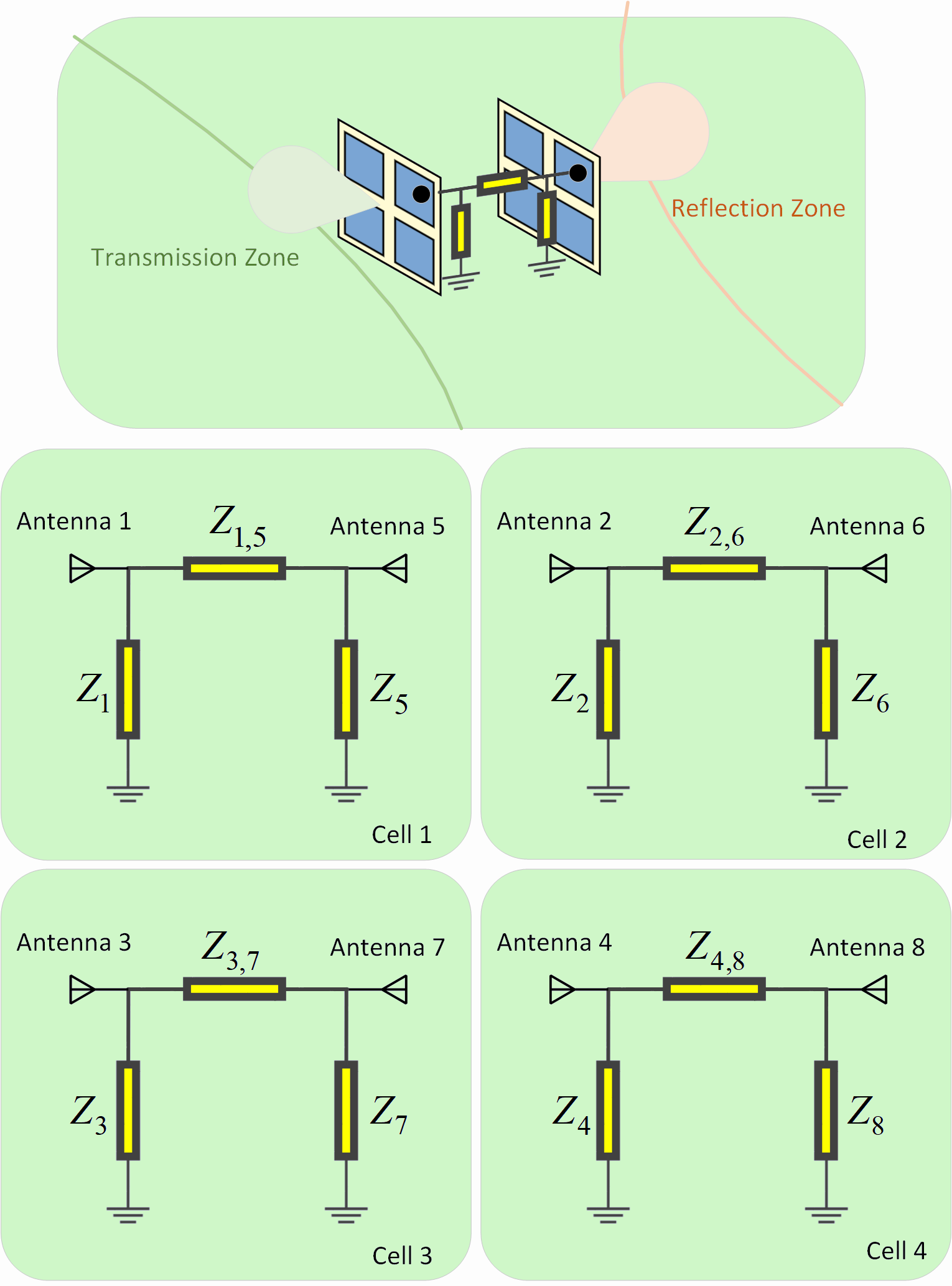}
\caption{STAR-BD-RIS with CW-SC architecture.}
\label{fig:star_bdris}
\end{figure}

In practical implementations, different impedance structures lead to different scattering matrix configurations. Here, a cell-wise single-connected (CW-SC) architecture is considered for the STAR-BD-RIS design, as shown in Fig.~\ref{fig:star_bdris}. In this structure, elements within each cell are interconnected through reconfigurable impedance components, while different cells remain isolated. Under the CW-SC configuration, each matrix ${{\mathbf{\Phi }}^{s}}$ becomes diagonal and can be written as~\cite{ref2,ref7,ref8}

\begin{equation}
{{\mathbf{\Phi }}^{s}}
=
\operatorname{diag}
\left(
{{\Phi }^{s,1}},
{{\Phi }^{s,2}},
\ldots ,
{{\Phi }^{s,F}}
\right),
\label{eq:diag_phi}
\end{equation}

where ${{\Phi }^{s,f}}\in \mathbb{C}$ represents the reflection coefficient of the $f$-th element in sector $s$. Accordingly, the power constraint for the STAR-BD-RIS is expressed as~\cite{ref2,ref7,ref8}

\begin{equation}
\sum\limits_{s}{|{{\Phi }^{s,f}}|^{2}}=1,
\quad \forall f.
\label{eq:power_constraint}
\end{equation}

The STAR-BD-RIS architecture provides a high degree of flexibility in controlling both reflection and transmission properties of the wireless environment~\cite{ref2,ref8}. This enables more efficient beamforming design, improved coverage, and enhanced system performance compared to conventional RIS, STAR-RIS, and BD-RIS architectures. Therefore, STAR-BD-RIS plays an important role in enabling intelligent and adaptive wireless environments for future 6G systems~\cite{ref2,ref7,ref8}.

% -------------------------------------------------
%      Advanced Multiple Access Techniques
% -------------------------------------------------
\section{Advanced Multiple Access Techniques}

Advanced multiple access techniques for next-generation wireless systems are presented in this section, with a focus on SCMA. These technologies are designed to improve spectral efficiency and support massive connectivity in dense wireless networks where a large number of users share limited spectral resources, as in 6G networks. Conventional orthogonal multiple access schemes allocate separate time or frequency resources to different users, which limits spectral efficiency in large-scale systems. To overcome this limitation, non-orthogonal and code-domain access techniques have been introduced, where multiple users can simultaneously share the same resources with controlled interference~\cite{ref1,ref5,ref21}.

% -------------------------------------------------
%                SCMA Structure
% -------------------------------------------------
\subsection{SCMA Structure}

\begin{figure}[!t]
\centering
\includegraphics[width=0.95\columnwidth]{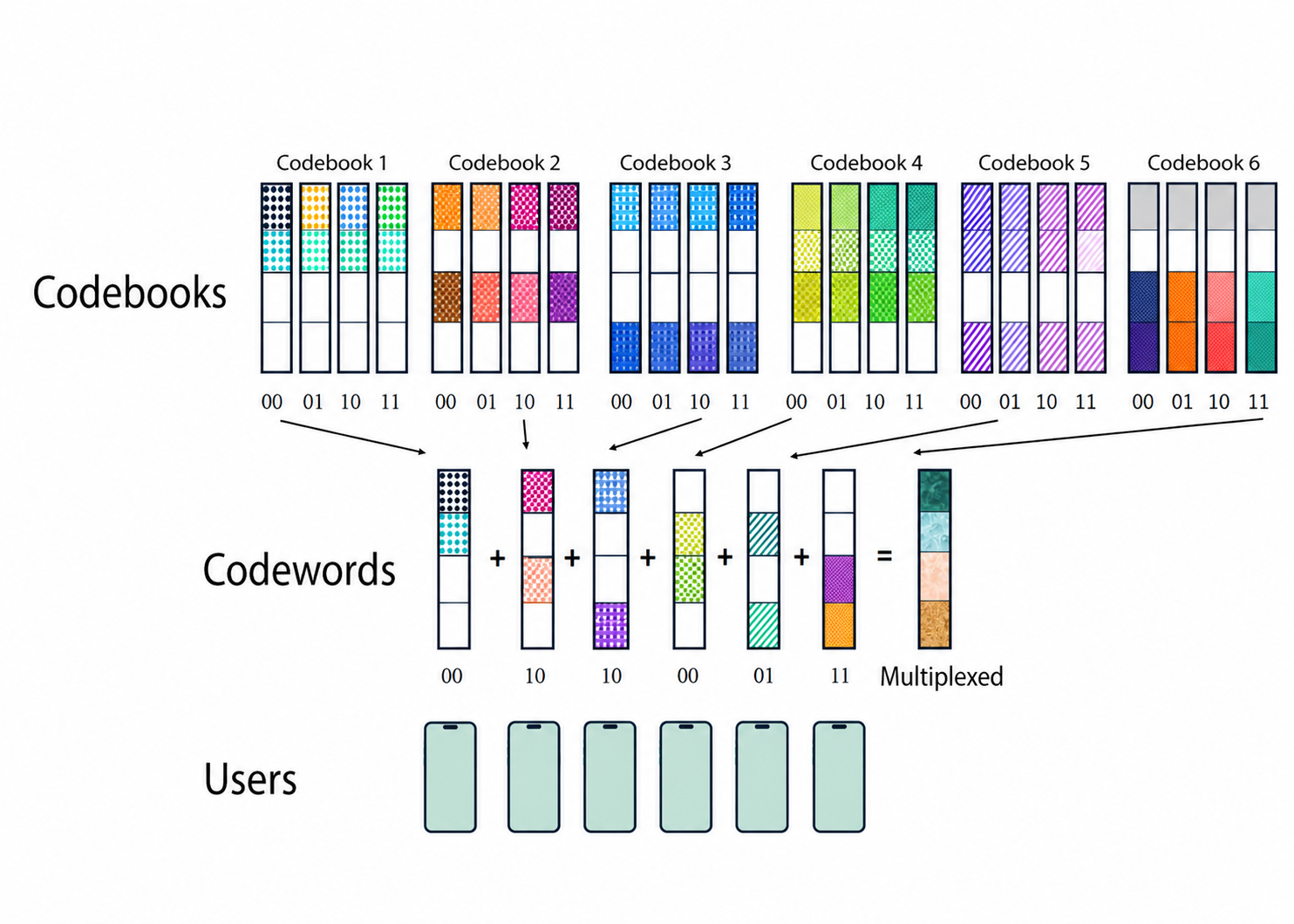}
\caption{SCMA codebooks and codeword structures.}
\label{fig:scma_structure}
\end{figure}

SCMA employs sparse multidimensional codebooks, where user information is directly mapped into sparse codewords. Figure~\ref{fig:scma_structure} shows SCMA codebooks and codeword structures. The relationship between users and subcarriers can be represented using a bipartite factor graph~\cite{ref1,ref5}. In the considered system, each user occupies ${{d}_{v}}$ subcarriers, while each subcarrier is shared by ${{d}_{f}}$ users. This sparse structure significantly reduces multi-user interference and enables efficient multi-user detection.

Let

\begin{equation}
\mathbf{X}
=
[
{{\mathbf{x}}_{1}},
{{\mathbf{x}}_{2}},
\ldots ,
{{\mathbf{x}}_{\mathbf{N}}}
]
\label{eq:scma_vector}
\end{equation}

denote the multi-user SCMA codeword vector. At the receiver, optimal detection can be formulated using the maximum a posteriori (MAP) criterion as~\cite{ref1}

\begin{equation}
\widehat{\mathbf{X}}
=
\underset{{{\mathbf{x}}_{{{n}_{s}}}},~\forall {{n}_{s}}}
{\operatorname{argmax}}
\,
\text{Pro}
(
\mathbf{X}
\mid
{{\mathbf{r}}_{\text{comm}}}
),
\label{eq:map_detection}
\end{equation}

where ${{\mathbf{r}}_{\text{comm}}}$ represents the received signal.

The detected symbol for user ${{n}_{s}}$ is obtained by marginalization as~\cite{ref5,ref21}

\begin{align}
&{{\widehat{\mathbf{x}}}_{{{n}_{s}}}}
=
\underset{{{\mathbf{x}}_{{{n}_{s}}}}}
{\operatorname{argmax}}
\,
\sum\limits_{\sim{{\mathbf{x}}_{{{n}_{s}}}}}
{
\left(
\text{Pro}(\mathbf{X})
\prod\limits_{c=1}^{C}
f
(
\mathbf{r}_{\text{comm}}^{c}
\mid
\mathbf{X}
)
\right)
},
\qquad \\ \nonumber
&{{n}_{s}}
=
\{1,\ldots ,\mathbf{N}\},
\label{eq:marginalization}
\end{align}
where $f(\cdot)$ is the conditional probability density function and $\text{Pro}(\mathbf{X})$ is the joint prior distribution.

Because brute-force MAP detection is exponentially complex, practical SCMA systems employ message passing algorithms (MPA). The computational complexity of MPA-based detection grows approximately as

\begin{equation}
{{S}^{{{d}_{f}}}},
\label{eq:complexity}
\end{equation}

where $S$ is the SCMA codebook size and ${{d}_{f}}$ is the number of users per subcarrier~\cite{ref1,ref5,ref21}.

Hence, SCMA provides an efficient framework for improving spectral efficiency and supporting massive connectivity. Its integration with advanced wireless architectures enables flexible interference management and enhanced system performance in next-generation wireless networks~\cite{ref1,ref21}.

% -------------------------------------------------
% Flexible Antenna and Intelligent Metasurface Technologies
% -------------------------------------------------
\section{Flexible Antenna and Intelligent Metasurface Technologies}

Flexible antenna and intelligent metasurface technologies play an important role in enabling adaptive and dynamic wireless environments for next-generation communication systems. Unlike conventional fixed antenna arrays, these technologies allow dynamic adjustment of antenna positions, structures, and electromagnetic responses, leading to improved channel conditions, enhanced beamforming capabilities, and higher system performance~\cite{ref7,ref8}.

Among these technologies, FIM has attracted significant attention due to its ability to dynamically reshape antenna structures. For instance, in a BS-equipped FIM system, the base station is equipped with an FIM-assisted antenna array composed of radiating elements indexed by $\mathcal{P}=\{1,\ldots ,P\}$. These elements are arranged as a uniform planar array on the $\text{y}$--$\text{z}$ plane, where the total number of elements is given by

\begin{equation}
P={{P}_{\text{y}}}{{P}_{\text{z}}}.
\label{eq:total_elements}
\end{equation}

Each antenna element can be dynamically adjusted along the $\text{x}$-axis, which enables spatial reconfiguration of the array. The position of the $p$-th element is defined as

\begin{equation}
{{\mathbf{o}}_{p}}
=
{{[{{x}_{p}},{{y}_{p}},{{z}_{p}}]}^{T}}
\in
{{\mathbb{R}}^{3}},
\quad
\forall p\in \mathcal{P}.
\label{eq:position_vector}
\end{equation}

The coordinates along the $\text{y}$- and $\text{z}$-axes are fixed, while the $\text{x}$-axis positions are adjustable. The overall array configuration is described by the morphing vector

\begin{equation}
\mathbf{x}
=
{{[{{x}_{1}},{{x}_{2}},\ldots ,{{x}_{P}}]}^{T}}.
\label{eq:morphing_vector}
\end{equation}

The displacement of each element is constrained by

\begin{equation}
0\le {{x}_{p}}\le {{x}_{\text{max}}},
\quad
\forall p\in \mathcal{P},
\label{eq:displacement_constraint}
\end{equation}

which ensures practical implementation feasibility~\cite{ref7,ref8}.

The channel between the base station and user ${{n}_{\text{s}}}$ depends on the FIM configuration and can be expressed as

\begin{equation}
{{\mathbf{g}}_{{{n}_{\text{s}}}}}(\mathbf{x})
=
\sum\limits_{d=1}^{D}
{{{\alpha }_{{{n}_{\text{s}}},d}}}
\mathbf{q}
(
\mathbf{x},
{{\varrho }_{d}},
{{\varphi }_{d}}
),
\label{eq:user_channel}
\end{equation}

where ${{\alpha }_{{{n}_{\text{s}}},d}}$ denotes the complex path gain, and $\mathbf{q}(\cdot)$ represents the array response vector given by

\begin{equation}
\mathbf{q}
(
\mathbf{x},
\varrho ,
\varphi
)
=
{{[
1,
\ldots ,
{{e}^{j\frac{2\pi }{\lambda }
(
{{x}_{p}}\sin \varphi \cos \varrho
+
{{y}_{p}}\sin \varphi \sin \varrho
+
{{z}_{p}}\cos \varphi
)}},
\ldots
]}^{T}}.
\label{eq:array_response}
\end{equation}

As illustrated in~\cite{ref7,ref8}, in a system model based on an FIM-equipped BS and STAR-BD-RIS, the channel between the base station and the intelligent surface is represented by

\begin{equation}
\mathbf{H}
=
\left[
{{\mathbf{h}}_{1}}(\mathbf{x}),
{{\mathbf{h}}_{2}}(\mathbf{x}),
\ldots ,
{{\mathbf{h}}_{F}}(\mathbf{x})
\right],
\label{eq:bs_ris_channel}
\end{equation}

where each component is given by

\begin{equation}
{{\mathbf{h}}_{f}}(\mathbf{x})
=
\sum\limits_{d=1}^{D}
{{{\alpha }_{f,d}}}
\mathbf{q}
(
\mathbf{x},
{{\varrho }_{d,f}},
{{\varphi }_{d,f}}
).
\label{eq:channel_component}
\end{equation}

The corresponding channel between the intelligent surface and the user is expressed as~\cite{ref8}

\begin{equation}
\mathbf{g}_{{{n}_{\text{s}}}}^{RIS}
=
{{\left[
{{k}_{{{n}_{\text{s}}},1}},
{{k}_{{{n}_{\text{s}}},2}},
\ldots ,
{{k}_{{{n}_{\text{s}}},F}}
\right]}^{T}}.
\label{eq:ris_user_channel}
\end{equation}

These flexible antenna technologies significantly improve system adaptability by dynamically adjusting antenna positions and electromagnetic responses. As a result, they enhance beamforming performance, increase channel diversity, and improve spectral efficiency in complex wireless environments~\cite{ref7,ref8}, which are essential for supporting dynamic scenarios and achieving efficient resource allocation in future communication systems.

% -------------------------------------------------
%            Resource Allocation Methods
% -------------------------------------------------
\section{Resource Allocation Methods}

This section reviews resource allocation methods for next-generation wireless communication systems, including classical optimization techniques and modern learning-based approaches. Efficient resource allocation is essential for managing power, bandwidth, beamforming, and other system parameters in complex and dynamic wireless environments.

Traditional resource allocation methods are mainly based on mathematical optimization techniques. These include convex optimization, fractional programming, and alternating optimization methods. In many cases, the resource allocation problem can be formulated as an optimization problem that aims to maximize system performance metrics such as spectral efficiency or energy efficiency, subject to various system constraints. A general form of such problems can be expressed as

\begin{equation}
{{\max }_{\mathbf{x}}}
\quad
f(\mathbf{x})
\quad
\text{s.t.}
\quad
\mathbf{x}\in \mathcal{X},
\label{eq:optimization_problem}
\end{equation}

where $\mathbf{x}$ represents the set of optimization variables, $f(\mathbf{x})$ denotes the objective function, and $\mathcal{X}$ is the feasible set defined by system constraints~\cite{ref1,ref2,ref3,ref4,ref5,ref6}.

Although classical optimization methods provide useful theoretical insights, they often suffer from high computational complexity, especially in large-scale systems with highly coupled variables. Moreover, these methods typically require accurate mathematical formulations and equations, and may not be suitable for real-time implementation in dynamic environments~\cite{ref1,ref21}.

To address these limitations, learning-based approaches have been introduced for resource allocation. In particular, DRL has emerged as a powerful tool for solving complex optimization problems in wireless networks. In DRL-based resource allocation, the problem is modeled as a Markov decision process, where an agent interacts with the environment to learn an optimal policy~\cite{ref1,ref2,ref3,ref4}.

At each time step $t$, the agent observes the system state ${{s}_{t}}$, selects an action ${{a}_{t}}$, and receives a reward ${{r}_{t}}$. The objective is to maximize the expected cumulative reward, which can be expressed as

\begin{equation}
{{\max }_{\pi }}
\quad
\mathbb{E}
\left[
\sum\limits_{t=0}^{\infty }
{{{\gamma }^{t}}}
{{r}_{t}}
\right],
\label{eq:drl_objective}
\end{equation}

where $\pi$ denotes the policy and $\gamma \in (0,1)$ is the discount factor~\cite{ref1,ref2,ref3,ref4}.

Among DRL algorithms, the deep deterministic policy gradient (DDPG)~\cite{ref1} method is widely used for continuous control problems such as power allocation and beamforming~\cite{ref1}. DDPG employs an actor-critic framework, where the actor network generates actions and the critic network evaluates their quality. The policy is updated based on the gradient of the expected return, which is given by

\begin{equation}
{{\nabla }_{\theta }}
J(\theta )
=
\mathbb{E}
\left[
{{\nabla }_{\theta }}
{{\pi }_{\theta }}(s)
{{\nabla }_{a}}
Q(s,a)
{{|}_{a={{\pi }_{\theta }}(s)}}
\right],
\label{eq:ddpg_gradient}
\end{equation}

where $\theta$ represents the parameters of the actor network~\cite{ref7}.

Despite their advantages, standard DRL methods often suffer from limited generalization capability and require extensive training when system conditions change. To overcome this issue, meta-learning approaches have been introduced. Meta-learning aims to enable models to learn how to learn, allowing them to quickly adapt to new environments with minimal additional training.

In meta-learning-based resource allocation, the model is trained over a distribution of tasks and learns a set of initial parameters that can be rapidly adapted. A commonly used formulation is

\begin{equation}
{{\min }_{\theta }}
\sum\limits_{i}
{{{\mathcal{L}}_{i}}}
\left(
\theta
-
\alpha
{{\nabla }_{\theta }}
{{\mathcal{L}}_{i}}(\theta )
\right),
\label{eq:meta_learning}
\end{equation}

where ${{\mathcal{L}}_{i}}$ denotes the loss function for task $i$, and $\alpha$ is the learning rate for task-specific adaptation~\cite{ref8}.

By combining meta-learning with DRL, it is possible to achieve fast adaptation and improved robustness in dynamic wireless environments~\cite{ref1,ref2,ref3,ref4,ref6,ref7,ref8}. These approaches reduce the need for retraining and enable efficient real-time decision-making. Overall, resource allocation methods based on learning techniques provide a flexible and scalable framework for handling complex optimization problems in next-generation wireless systems. Their integration with advanced communication technologies is essential for achieving intelligent and adaptive network operation.

% -------------------------------------------------
% Open Research Challenges and Future Directions
% -------------------------------------------------
\section{Open Research Challenges and Future Directions}

Despite the significant progress in integrating ISAC, RIS-based architectures, advanced multiple access schemes, flexible antenna technologies, and learning-based resource allocation methods, several important challenges remain open~\cite{ref1,ref22}. Addressing these issues is essential for the practical realization of next-generation wireless networks.

% -------------------------------------------------
% Joint System Integration and Unified Design
% -------------------------------------------------
\subsection{Joint System Integration and Unified Design}

One of the main challenges lies in the joint integration of multiple emerging technologies within a unified framework. Combining ISAC, STAR-BD-RIS, SCMA, and FIM-based architectures leads to highly coupled system models with a large number of optimization variables. Designing a unified optimization framework that can efficiently handle beamforming, phase-shift design, antenna configuration, and multiple access simultaneously remains a complex problem. Future research should focus on developing scalable and structured frameworks that can jointly optimize these components with manageable complexity.

% -------------------------------------------------
% Scalability and Computational Complexity
% -------------------------------------------------
\subsection{Scalability and Computational Complexity}

As the number of users, antennas, and intelligent surface elements increases, the dimensionality of the optimization problem grows significantly. This results in high computational complexity, which limits the applicability of conventional optimization methods and even some learning-based approaches. Although DRL and meta-learning methods reduce computational burden during online operation, their training process can still be expensive. Therefore, designing low-complexity and scalable algorithms, possibly through model reduction, distributed learning, or hierarchical optimization, is an important research direction.

% -------------------------------------------------
% Real-Time Implementation and Practical Constraints
% -------------------------------------------------
\subsection{Real-Time Implementation and Practical Constraints}

Real-time implementation is a critical requirement in practical wireless systems. Most existing works assume ideal conditions, such as perfect CSI and unlimited computational resources. However, in real scenarios, CSI is often imperfect, and hardware limitations must be considered. Moreover, latency constraints in applications such as autonomous systems and IoT require fast decision-making. Future work should consider practical constraints, including imperfect CSI and hardware impairments, to develop more realistic and implementable solutions.

% -------------------------------------------------
% Robustness and Generalization
% -------------------------------------------------
\subsection{Robustness and Generalization of Learning-Based Methods}

Although DRL and meta-learning approaches have shown promising performance, their robustness and generalization capability remain challenging issues. Learning-based models may suffer performance degradation when the environment changes significantly from the training conditions. Meta-learning partially addresses this issue by enabling faster adaptation; however, further improvements are still needed. Future research should focus on robust learning frameworks that can generalize well across different scenarios, possibly by incorporating uncertainty modeling, transfer learning, and online adaptation mechanisms.

% -------------------------------------------------
% Energy Efficiency and Sustainable Design
% -------------------------------------------------
\subsection{Energy Efficiency and Sustainable Design}

Energy consumption is a key concern in next-generation wireless networks, particularly with the deployment of large-scale antenna arrays and intelligent surfaces. While RIS technologies reduce the need for active components, the overall system still requires efficient energy management. Learning-based methods can help optimize energy efficiency; however, their training process may consume significant computational power. Future work should consider energy-aware algorithm design, green communication strategies, and hardware-efficient implementations.

% -------------------------------------------------
% Hardware Implementation and Standardization
% -------------------------------------------------
\subsection{Hardware Implementation and Standardization}

Another important challenge is the practical implementation of advanced technologies such as STAR-BD-RIS and FIM. Issues such as hardware complexity, cost, reliability, and compatibility with existing communication standards must be addressed. In addition, there is currently a lack of unified standards for integrating these technologies into real-world systems. Future research should focus on hardware prototyping, experimental validation, and standardization efforts to facilitate the deployment of these technologies in commercial networks.

% -------------------------------------------------
% Security and Privacy Considerations
% -------------------------------------------------
\subsection{Security and Privacy Considerations}

With the increasing use of intelligent and adaptive wireless systems, security and privacy concerns become more critical. RIS-assisted systems and learning-based algorithms may introduce new vulnerabilities, such as eavesdropping, adversarial attacks, and data leakage. Ensuring secure communication and protecting user data in such complex environments is an open challenge. Future directions include the development of secure beamforming techniques, robust learning algorithms, and privacy-preserving optimization methods.

% -------------------------------------------------
%                   Conclusion
% -------------------------------------------------
\section{Conclusion}

In this paper, we presented a comprehensive review of emerging technologies and advanced resource allocation strategies for next-generation wireless communication systems. Specifically, we investigated ISAC, RIS-assisted architectures including STAR-RIS and STAR-BD-RIS, advanced multiple access schemes such as SCMA, as well as flexible antenna and intelligent metasurface technologies. These technologies enable highly adaptive, efficient, and intelligent wireless environments that are essential for future 6G networks.

We first analyzed the fundamental principles and system models of ISAC, highlighting its capability to unify sensing and communication functionalities within a shared framework. The inherent tradeoff between sensing accuracy and communication performance was discussed, emphasizing the need for joint optimization strategies. Subsequently, RIS-assisted systems were examined, where advanced architectures such as STAR-BD-RIS were shown to provide enhanced flexibility in controlling electromagnetic wave propagation, thereby significantly improving spectral efficiency, coverage, and interference management.

Furthermore, advanced multiple access techniques, including SCMA, were reviewed as key enablers for supporting massive connectivity and improving spectral efficiency in dense wireless networks. Their ability to efficiently manage interference and utilize limited resources makes them highly suitable for next-generation systems. In addition, flexible antenna technologies such as fluid antenna systems were discussed, demonstrating their potential to dynamically adapt to changing channel conditions and enhance system performance.

From a resource allocation perspective, we reviewed both classical optimization methods and modern learning-based approaches. While traditional techniques provide valuable theoretical insights, they often suffer from high computational complexity and limited adaptability. In contrast, DRL and meta-learning-based approaches offer scalable and intelligent solutions capable of handling complex and dynamic environments. In particular, meta-learning enhances the generalization and adaptability of DRL algorithms, enabling fast convergence and efficient real-time decision-making.

Finally, we identified several open research challenges and future directions, including joint system integration, scalability, real-time implementation, robustness of learning-based methods, energy efficiency, hardware constraints, and security considerations. Addressing these challenges is crucial for the practical deployment of next-generation wireless systems. Future research efforts should focus on developing unified frameworks and practical implementations to fully realize the potential of these technologies in real-world scenarios.

% -------------------------------------------------
%                 List of Symbols
% -------------------------------------------------
\section*{List of Symbols}

The list of symbols is presented in Table~\ref{tab:symbols}.

\begin{table*}[!t]
\centering
\caption{List of Symbols}
\label{tab:symbols}
\renewcommand{\arraystretch}{1.2}
\begin{tabular}{|c|c|c|c|}
\hline
\textbf{Symbol} & \textbf{Description} & \textbf{Symbol} & \textbf{Description} \\
\hline

${{a}_{t}}$ & Action & ${{a}_{u}}$ & Complex path gain of $u$-th target \\
\hline

$C$ & Number of subcarriers & $d({{\theta }_{l}})$ & Desired beampattern \\
\hline

${{d}_{f}}$ & Users per subcarrier & ${{d}_{v}}$ & Subcarriers per user \\
\hline

$f(\cdot )$ & Conditional PDF & $f(\mathbf{x})$ & Objective function \\
\hline

$J(\theta )$ & Expected return & ${{k}_{{{n}_{\text{s}}},f}}$ & Channel coefficient \\
\hline

$L$ & Number of samples & $\lambda$ & Wavelength \\
\hline

${{\mathcal{L}}_{i}}$ & Loss function & $\mathcal{P}$ & Antenna index set \\
\hline

$\mathcal{S}$ & Set of sectors & $\mathcal{X}$ & Feasible set \\
\hline

$\pi$ & Policy & ${{\phi }_{u}}$ & Target angle \\
\hline

${{\Phi }^{s,f}}$ & Reflection coefficient & $\mathbf{g}_{{n}_{\text{s}}}$ & BS-user channel \\
\hline

$\mathbf{g}_{{n}_{\text{s}}}^{RIS}$ & RIS-user channel & ${{\mathbf{h}}_{f}}$ & Channel component \\
\hline

$\mathbf{H}$ & Channel matrix / BS-RIS channel & ${{\mathbf{I}}_{F}}$ & Identity matrix \\
\hline

$\mathbf{n}$ & Additive noise vector & $\mathbf{N}$ & Radar noise matrix \\
\hline

${{\mathbf{o}}_{p}}$ & Position vector & $\mathbf{q}(\cdot )$ & Array response \\
\hline

${{\mathbf{r}}_{\text{comm}}}$ & Received signal & $\mathbf{R}$ & Covariance matrix \\
\hline

$\mathbf{s}$ & Information symbols & $\mathbf{W}$ & Beamforming matrix \\
\hline

$\mathbf{x}$ & Transmitted signal / Morphing vector & ${{\mathbf{x}}_{u}}$ & User codeword \\
\hline

$\mathbf{X}$ & SCMA codeword matrix & $\mathbf{y}$ & Received signal \\
\hline

${{\mathbf{Y}}_{\text{radar}}}$ & Radar echo signal & ${{\widehat{\mathbf{x}}}_{u}}$ & Estimated symbol \\
\hline

$\widehat{\mathbf{X}}$ & Estimated codeword & $\mathbf{\Phi}$ & STAR-BD-RIS matrix \\
\hline

${{\mathbf{\Phi }}^{s}}$ & Sector matrix & $\alpha$ & Learning rate \\
\hline

${{\alpha }_{{{n}_{\text{s}}},d}}$ & Path gain & ${{\alpha }_{r}}(\phi )$ & Receive steering vector \\
\hline

${{\alpha }_{t}}(\phi )$ & Transmit steering vector & ${{\epsilon }_{\text{Beam}}}$ & Beam error \\
\hline

$\gamma$ & Discount factor & $P$ & Number of antennas \\
\hline

$P({{\theta }_{l}})$ & Generated beampattern & $Q(s,a)$ & Q-function \\
\hline

${{r}_{t}}$ & Reward & ${{s}_{t}}$ & State \\
\hline

$S$ & Codebook size & $U$ & Number of users/targets \\
\hline

${{x}_{\text{max}}}$ & Maximum displacement & ${{x}_{p}},{{y}_{p}},{{z}_{p}}$ & Coordinates \\
\hline

$\varrho ,\varphi$ & Angles & $\theta$ & Actor parameters \\
\hline

\end{tabular}
\end{table*}

% -------------------------------------------------
%                   References
% -------------------------------------------------

\end{document}